\documentclass[referee]{aa} % for a referee version

\usepackage{graphicx}
\usepackage{natbib}
\usepackage{latexsym}
\usepackage{amsmath}
\usepackage{amssymb}
\usepackage{amstext}
\usepackage{color}
\usepackage{psfrag}
\usepackage{txfonts}
\usepackage{array} % con este paquete uso la opción \setlength{\extrarowheight}{3pt} en la tabla
\def\km{km s$^{-1}$}

\def\d{$^\circ$}
\def\m{$^\prime$}
\def\s{$^{\prime\prime}$}

\def\ca{cm$^{-3}$}
\def\cm2{cm$^{-2}$}
\def\pp{$^{\prime\prime}$}
\def\2{$^{12}$CO}
\def\3{$^{13}$CO}
\def\msol{M$_\odot$}

\begin{document}

\title{An X-ray study of the supernova remnant G20.0-0.2 and its surroundings}
\author {A. Petriella \inst{1,2}
\and S. A. Paron \inst{1,2,3}
\and E. B. Giacani  \inst{1,3}
}

\institute{Instituto de Astronom\'{\i}a y F\'{\i}sica del Espacio (CONICET-UBA),
             CC 67, Suc. 28, 1428 Buenos Aires, Argentina\\
            \email{apetriella@iafe.uba.ar}
\and CBC - Universidad de Buenos Aires, Argentina
\and FADU - Universidad de Buenos Aires, Argentina
}

\date{Received <date>; Accepted <date>}

\abstract
{}
{We study the supernova remnant G20.0-0.2 and its surroundings in order to look for the high energy counterpart 
of the radio nebula and to find evidence of interaction between the shock front and the interstellar medium. }
{We used Chandra archival observations to analyze the X-ray emission from the supernova remnant.
The surrounding gas was investigated using data extracted from the Galactic Ring Survey, the VLA Galactic Plane Survey,
the Galactic Legacy Infrared Midplane Survey Extraordinaire, and the Bolocam Galactic Plane Survey.}
{G20.0-0.2 shows diffuse X-ray emission from the central region of the radio remnant. Although the current data do not allow 
us to distinguish between a thermal or non-thermal origin for the X-ray diffuse emission, based on the radio properties we
suggest a synchrotron origin as the most favorable.
The hard X-ray point source CXO J182807.4-113516 appears located at the geometrical center of the remnant and 
is a potential candidate to be the pulsar powering the nebula. 
We found a molecular cloud adjacent to the flattest border of G20.0-0.2, indicating a probable 
interaction between the shock front of the remnant and the molecular gas. 
Several young stellar object candidates are found located in the brightest region of the molecular emission, and over a 
millimeter continuum source and a dark cloud. This distribution is an indication of an active star forming region around 
the supernova remnant. }
{}

\titlerunning{An X-ray study of the SNR G20.0-0.2 and its surroundings}
\authorrunning{A. Petriella et al.}

\maketitle
   
\section{Introduction}
\label{present}

\begin{figure}[ht]
\centering
\includegraphics[width=8cm]{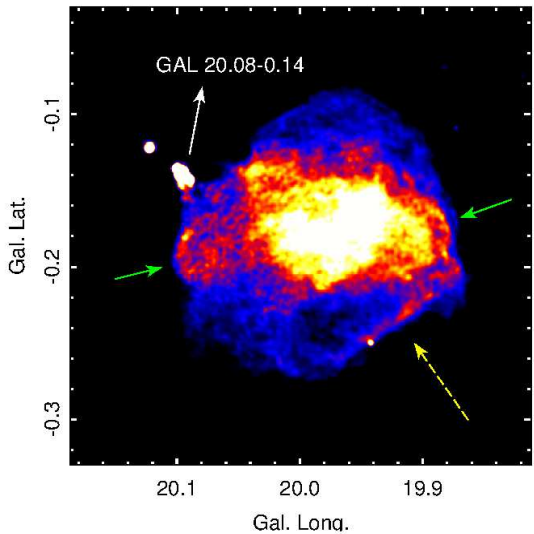}
\caption{Radio continuum emission of G20.0-0.2 at 20 cm. We marked the position of the complex of UCHII regions GAL 20.08-0.14. 
The green arrows mark the arc-like filaments and the yellow arrow indicates the flattest border and the radio filament.}
\label{fig_radio}
\end{figure}

The \object{SNR G20.0-0.2} (hereafter G20) was classified as a plerion\footnote{Plerions are also referred as pulsar wind nebulae (PWNe).}
by \citet{becker85} based on the filled-center morphology, 
the presence of significant polarization at 6 cm, and a flat radio spectral index $\alpha\sim0.0$ ($S_{\nu}\propto\nu^{\alpha}$).
In Fig. \ref{fig_radio}, we show the best image of the radio continuum emission of the SNR at 20 cm, 
extracted from the MAGPIS \citep{helfand06}. The radio synchrotron emission from G20 has a complex
morphology, with multiple features like bright knots, arcs and filaments.
The overall emission is dominated by an elliptical central core of about 3\farcm8~$\times$~2\farcm2, with the major axis oriented
in the direction of the Galactic plane. This bright feature is surrounded by faint emission.
Two arc-like filaments are located near the outer edges of G20 (green arrows in the figure).
Interestingly, one border of the SNR is clearly flat and is delineated by a bright radio filament (yellow arrow),
suggesting that it may have encountered a higher ambient density in this direction.

PWNe with complex morphologies have been observed in several cases, such as G328.4+0.2 \citep{gelfand07}
and G0.9+0.1 \citep{dubner08}, but the presence of a flat radio filament is a common characteristic
of a surrounding radio shell, with an expected steeper radio spectrum ($\alpha<-0.3$).
However, the other available observations at 330 MHz do not have an adequate angular resolution to
look for variations in the radio spectral index through the remnant and confirm the presence of a radio shell.   

To the north of G20, lies the complex of ultra-compact HII regions \object{GAL 20.08-0.14} \citep{wood89}, 
which appears as a compact and bright radio source adjacent to the SNR. Different masers and molecular lines 
have been detected toward the HII regions \citep{avedisova02Cat,galvan09}, with velocities between $\sim$40 and $\sim$45 \km~(all
velocities are referred to the local standard of rest, hereafter LSR).
\citet{anderson09a} reported the presence of a molecular cloud (MC) that spatially coincides with GAL 20.08-0.14 at a velocity 
of $\sim$42~\km. From an analysis of the HI absorption spectrum, they concluded that the complex of HII regions
lies at its far distance, namely $\sim$12.6 kpc.         

In the X-ray domain, the remnant remains almost unexplored. \citet{hands02} performed a X-ray survey of selected 
regions of the Galactic plane with the {\it XMM-Newton} satellite and reported the presence of some 
diffuse X-ray emission in direction to G20. 
Additionally, the SNR is the only source that lies partially inside the error box of the Fermi-LAT $\gamma$-ray source 
J1828.3-1124c \citep{nolan12}. For this reason, the authors suggested a connection between 
the remnant and the $\gamma$-ray emission.

In this work, we present the first X-ray study of the SNR G20.0-0.2 using {\it Chandra} observations 
with the aim to establish if this remnant is purely plerionic or a composite one.
We also search for sources candidate to be the pulsar that is powering the emission.
In addition, we investigate the interstellar medium (ISM) around G20 looking for signs of 
interaction with the expanding shock of the SNR, which could explain the peculiar 
morphology that this remnant shows in the radio band. 

\section{Results and discussion}
\label{res}

\subsection{X-ray emission from G20.0-0.2}
\label{secc_xray}

We reprocessed an archival {\it Chandra} observation of G20 obtained on 2005 November 5 (ObsID: 5563) 
using ACIS-I in \texttt{vfaint} mode.
The data were calibrated with CIAO (version 4.4) and CALDB (version 4.4.8), provided by the Chandra X-ray Center (CXC).
After filtering the periods of high count-rate, we produced a cleaned event file with 34.45 ks of observation, 
which was used to perform an imaging and spectral analysis.  

\subsubsection{Imaging}
\label{imagx}

\begin{figure}[ht]
\centering
\includegraphics[width=7.5cm]{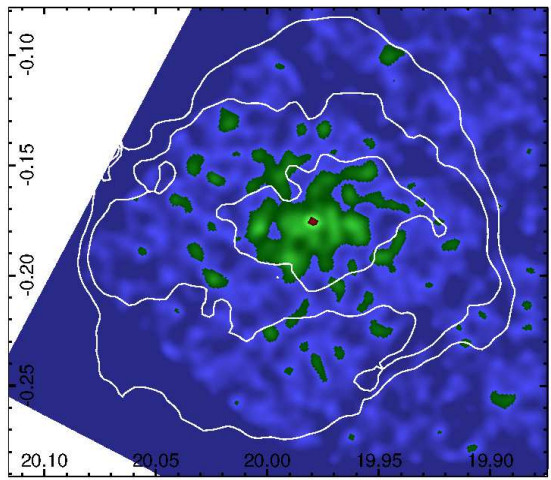}
\caption{Exposure corrected X-ray image between 2.0 and 7.0 keV, in Galactic coordinates.
Point sources outside the diffuse emission have been subtracted following the procedure described in the text.
The contours are the radio continuum emission of the SNR at 20 cm (contour levels are 1.0, 1.7, and 3.0 mJy/beam).}
\label{figX1a}
\end{figure}

We analyzed the event file and found an enhancement of the X-ray emission toward the central radio core of G20.
We constructed exposure corrected images in different energy bands and found diffuse emission above 2 keV and a
single point source embedded on it. Additionally, we noticed several X-ray clumps outside this extended emission. 
Most of them are likely point sources that appear broadened by the smoothing. 
To emphasize the diffuse emission, we produced a point sources subtracted image following
the standard procedure described in the Ciao Science Threads. We used the tool \texttt{wavdetect} to detect the point source
candidates and we visually inspected the result to exclude those sources that appear to be actual diffuse emission.
For each source, we created the source and background regions with the tool \texttt{roi} and replaced the source region with
the mean background region using \texttt{dmfilth}. To increase the signal-to-noise ratio of the image, we convolved 
it with a Gaussian kernel. The final image has a spatial scale of 0.05 arcmin/pixel.
Fig. \ref{figX1a} shows the point sources subtracted X-ray image of the SNR between 2.0 and 7.0 keV, 
overlayed with some contours of the radio continuum emission.
We noticed that the diffuse emission is confined within an 
ellipse of 1\farcm8~$\times$~1\farcm1, which is enclosed by the central bright radio core and is elongated in the same direction
as in the radio band.

\begin{figure}[ht]
\centering
\includegraphics[width=7.5cm]{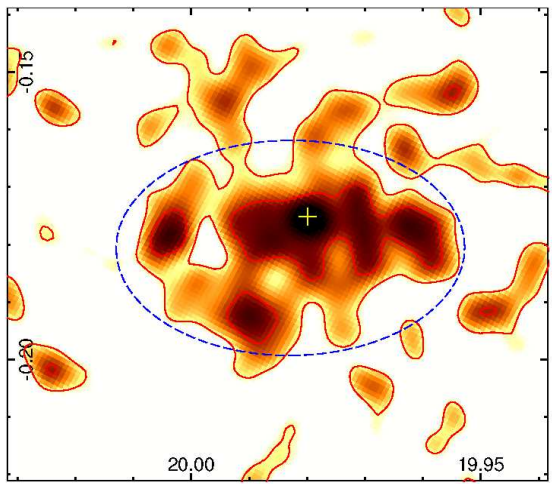}
\caption{Exposure corrected X-ray image between 2.0 and 7.0 keV, enlarged toward the diffuse emission.  
The yellow cross is CXO J182807.4-113516. The ellipse indicates the region used to extract the spectrum.}
\label{figX1b}
\end{figure}

Fig. \ref{figX1b} shows an X-ray image in the 2.0-7.0 keV energy band toward the diffuse emission. 
Interestingly, we noted the presence of only one X-ray point source embedded in the diffuse 
X-ray emission (indicated in the figure with a yellow plus sign). 
This source has been identified in the Chandra Source Catalog (CSC, \citealt{evans10})
as \object{CXO J182807.4-113516} and is centered at R.A. = 18$^{h}$28$^{m}$7$^{s}$.43, Dec. = -11\d35\m16\farcs32,
almost at the geometrical center of the radio remnant. 
In Sect. \ref{spectrum}, we will discuss about its nature and the connection with the remnant.

\subsubsection{X-ray spectral analysis}
\label{spectrum}

To perform a spectral analysis of the diffuse emission from the SNR, we extracted the spectrum with the 
CIAO task \texttt{specextract} from the elliptical region indicated in Fig. \ref{figX1b}. 
This region excludes some of the diffuse emission 
but reduces the contamination from the background. The point source CXO J182807.4-113516 was removed from the extraction region.  
For the background, we selected a circular region free of diffuse emission and point sources. 
The spectrum was binned with a minimum of 15 counts/bin. 
We extracted a total of 1059 counts between 1.0 and 7.0 keV. We fitted the spectrum using XSPEC (version 12.7.1) and $\chi^{2}$ statistics.
%The unbinned spectrum and {\it cash} statistics was used to derive the model parameters. 

\begin{figure}[ht]
\centering
\includegraphics[width=6cm,angle=-90]{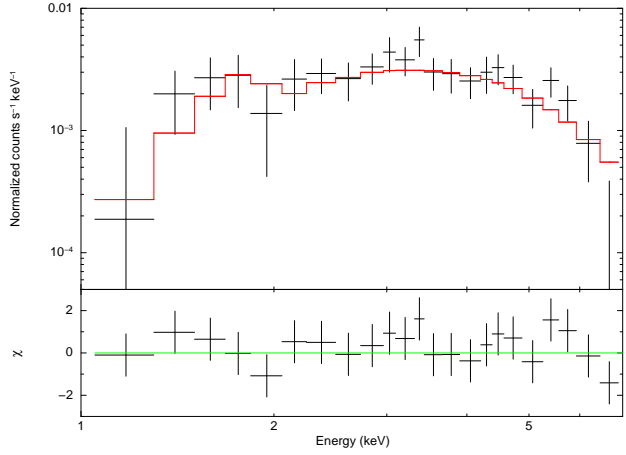}
\caption{Chandra ACIS-I spectrum for the diffuse emission from G20.0-0.2. The solid line is the best-fit absorbed power-law 
model with a hydrogen column density fixed at 4$\times10^{22}$~\cm2.}
\label{figX2}
\end{figure}

The Chandra ACIS-I spectrum of G20 presents a continuum without evidence of emission lines (Fig. \ref{figX2}).
We fitted it with both non-thermal and thermal models. 
To probe a possible synchrotron origin, we fitted the spectrum with an absorbed {\it power-law}. 
For the absorption model ({\it wabs}), we fixed the hydrogen column density to the value N$_{H}\sim4\times10^{22}$~\cm2,
obtained through the analysis of the neutral and molecular gas (Sect. \ref{secc_molec1}). 
The best fit yields a photon index $\Gamma_{X}$ of 1.87, a value similar to that observed in other PWNe \citep{kargal10}.
Letting the absorption to vary freely does not produce any significant improvement in the fitting and 
results in N$_{H}\sim3.4\times10^{22}$~\cm2, a value that is still comparable to the one measured using the neutral and molecular data. 

\begin{table*}[ht]
\caption{Best-fit parameters for the diffuse X-ray emission of G20.0-0.2 between 1.0 and 7.0 keV. 
The symbol (*) indicates that the parameter was frozen. For the {\it nei} model, the abundance is solar (from \citealt{anders89}). 
$F$ is the unabsorbed flux. The unabsorbed luminosity $L$ was calculated for a distance of 4.5 kpc (see Sect. \ref{secc_molec1}). 
Errors quoted are 90\%.}
\label{table2}
%\tiny
\centering
\setlength{\extrarowheight}{3pt}
\begin{tabular}{ccccccc}
\hline\hline
Model    & $\chi^{2}/{d.o.f.}$ & N$_{H}$ & $\Gamma_{X}$ & T$_{e}$ & F (1.0-7.0 keV)                            &  L (1.0-7.0 keV)    \\
         &      & [$\times10^{22}$ cm$^{-2}$]  &  & [keV] &[$\times10^{-13}$ erg cm$^{-2}$ s$^{-1}$ ] &  [$\times10^{33}$ erg s$^{-1}$] \\
\hline
power-law & 62.16/62 & 4*                     & 1.87$_{-0.44}^{+0.41}$ & -         & 5.29$_{-0.85}^{+0.93}$ & 1.28$_{-0.20}^{+0.23}$ \\
power-law & 61.97/61 & 3.41$_{-1.65}^{+2.70}$ & 1.64$_{-0.83}^{+1.07}$ & -         & 4.63$_{-1.37}^{+4.75}$ & 1.12$_{-0.88}^{+1.15}$ \\
nei      & 64.84/61 & 4*                     & -              & 4.97$_{-1.79}^{+4.06}$ & 5.11$_{-0.71}^{+0.71}$ & 1.24$_{-0.17}^{+0.17}$ \\
nei      & 64.83/60 & 4.1$_{-1.35}^{+2.41}$  & -              & 4.82$_{-2.40}^{+7.01}$ & 5.51$_{-1.62}^{+4.08}$ & 1.34$_{-0.39}^{+0.98}$ \\
%\hspace{1 pt}
\hline
\end{tabular}
%\tablefoot{2MASS Qual.: }
\end{table*}

Regarding the thermal models, a thermal plasma in collisional equilibrium ({\it apec}) 
requires a considerably high electron temperature for a SNR (T$_{e}\sim$19~keV) when the abundance parameter is kept 
equal to the solar value of \citet{anders89}, and a temperature of $\sim$7~keV when the abundance is a free parameter. 
In this last case, although the temperature is not extremely high, the best fit abundance yields an unrealistic low value ($\sim$10$^{-9}$). 
For a non-equilibrium plasma ({\it nei}), freezing the abundance to the solar value results in T$_{e}\sim$5~keV, which approaches
the expected temperature of a young SNR \citep{vink12}. The obtained ionization parameter ($\tau < 10^{-9}$~cm$^{-3}$s) also
points to a plasma out of equilibrium. In Table \ref{table2}, we list the results of the 
spectral fitting for the {\it power-law} and the {\it nei} models.

In conclusion, the spectral fits themselves do not preclude a thermal origin for the X-ray emission from G20. However, based
on the radio properties, we propose that a synchrotron origin for the high energy emission is the most probable. 
The mere presence of a PWN component indicates that there is a compact source powering it. 
The best candidate to be a neutron star is CXO J182807.4-113516, the only X-ray point source embedded in the diffuse emission and 
located at the geometrical center of the SNR. 
There is no radio counterpart in the VLA image, nor there is any optical or infrared source within 
the position error box in the USNOb1.0, 2MASS\footnote{The Two Micron All Sky Survey (2MASS) is a joint project of the University of Massachusetts and the Infrared Processing and Analysis Center/California Institute of Technology, funded by the National Aeronautics and Space Administration and the National Science Foundation.}, and Spitzer catalogs.
The closest point sources are USNOb1.0 0784-0429139, 2MASS 18280728-1135224, 
and Spitzer G019.9794-00.1736, located at $\sim$9\s, $\sim$6\s, and $\sim$6\s~from CXO J182807.4-113516, respectively.
These distances are larger than three times the combined X-ray and optical/infrared positional error. 

Although the low number of counts ($<$20) from CXO J182807.4-113516 does not allow to perform a spectral 
fitting or timing analysis to investigate its nature, we can still use the poor spectral information to get a rough picture of this source.  
We found that all the counts from CXO J182807.4-113516 originate between 3.0 and 8.0 keV, so we defined
the hardness ratio $H$ as the ratio of the counts in the 5.0-8.0 keV band to the counts in the 3.0-5.0 keV band.    
We extracted the net (i.e., background subtracted) number of counts and obtained 
11.82$\pm$3.75 counts between 3.0 and 5.0 keV and 6.33$\pm$2.73 counts between 5.0 and 8.0 keV, 
which yields to H=0.54$\pm$0.29.
This value can be compared with the predicted value of $H$ estimated by assuming different emission models. 
The X-ray emission from pulsars (PSRs) may correspond either to blackbody emission from the neutron star surface 
or to non-thermal emission from the magnetosphere. 
Using the tool W3PIMMS\footnote{http://heasarc.gsfc.nasa.gov/Tools/w3pimms.html} and the net count rate 
between 3.0 and 8.0 keV (0.00053~cts/s), we modeled the X-ray emission from CXO J182807.4-113516. 
We fixed the hydrogen column density to $4\times10^{22}$ cm$^{-2}$, to match the absorption of the diffuse emission. 
For a blackbody with a temperature of 0.1 keV, the predicted $H$ is $\sim$8$\times10^{-9}$, in complete 
disagreement with the measured value. 
For a hot blackbody (1 keV), we obtained $H\sim0.28$, but such high temperature is not
expected in a young PSR. 
According to \citet{kargal10}, the non-thermal spectrum of a PSR powering a PWN is well described by a power-law model 
with photon index in the range $1.0\lesssim\Gamma_{X}\lesssim2.0$. 
The predicted values of $H$ are $\sim$0.59 (for $\Gamma_{X}=1.0$) and $\sim$0.39 (for $\Gamma_{X}=2.0$), in good
agreement with the measured value of H=0.54$\pm$0.29.
Thus, we conclude that the emission from CXO J182807.4-113516 is more likely non-thermal in origin, although 
a deeper X-ray exposure of this source is needed to confirm its nature.
Using W3PIMM, we estimated its luminosity in the 0.5-8.0 keV band for a distance of 4.5 kpc, 
obtaining $\sim$0.7$\times10^{32}$ erg/s (for $\Gamma_{X}=1.0$) and $\sim$1.2$\times10^{32}$ erg/s (for $\Gamma_{X}=2.0$).
From the best fit power-law model, the X-ray luminosity of G20 between 0.5 and 8.0 keV is $\sim$1.8$\times10^{33}$ erg/s.
The luminosities of G20 and CXO J182807.4-113516 are in good agreement with those reported by \citet{kargal10} for
other PWNe and their powering PSRs.
Thus, based on its position and spectral behavior, CXO J182807.4-113516 appears as a good candidate to be the pulsar
created after the explosion of the supernova that originated G20 and that is powering the synchrotron nebula.

\subsection{The interstellar medium around G20.0-0.2}
\label{secc_molec1}

We studied the ISM around G20 to investigate the presence of
molecular clouds (MCs) that may have affected the expansion of the SNR shock front. 
The molecular data were extracted from the Galactic Ring Survey (GRS,
\citealt{jackson06}). The survey maps the Galactic ring in the $^{13}$CO J=1--0
line with angular and spectral resolutions of 46\arcsec~and 0.2 km s$^{-1}$, respectively. The
observations were performed in both position-switching and on-the-fly mapping modes, achieving an
angular sampling of 22\arcsec.

\begin{figure*}[ht]
\centering
\includegraphics[width=14cm]{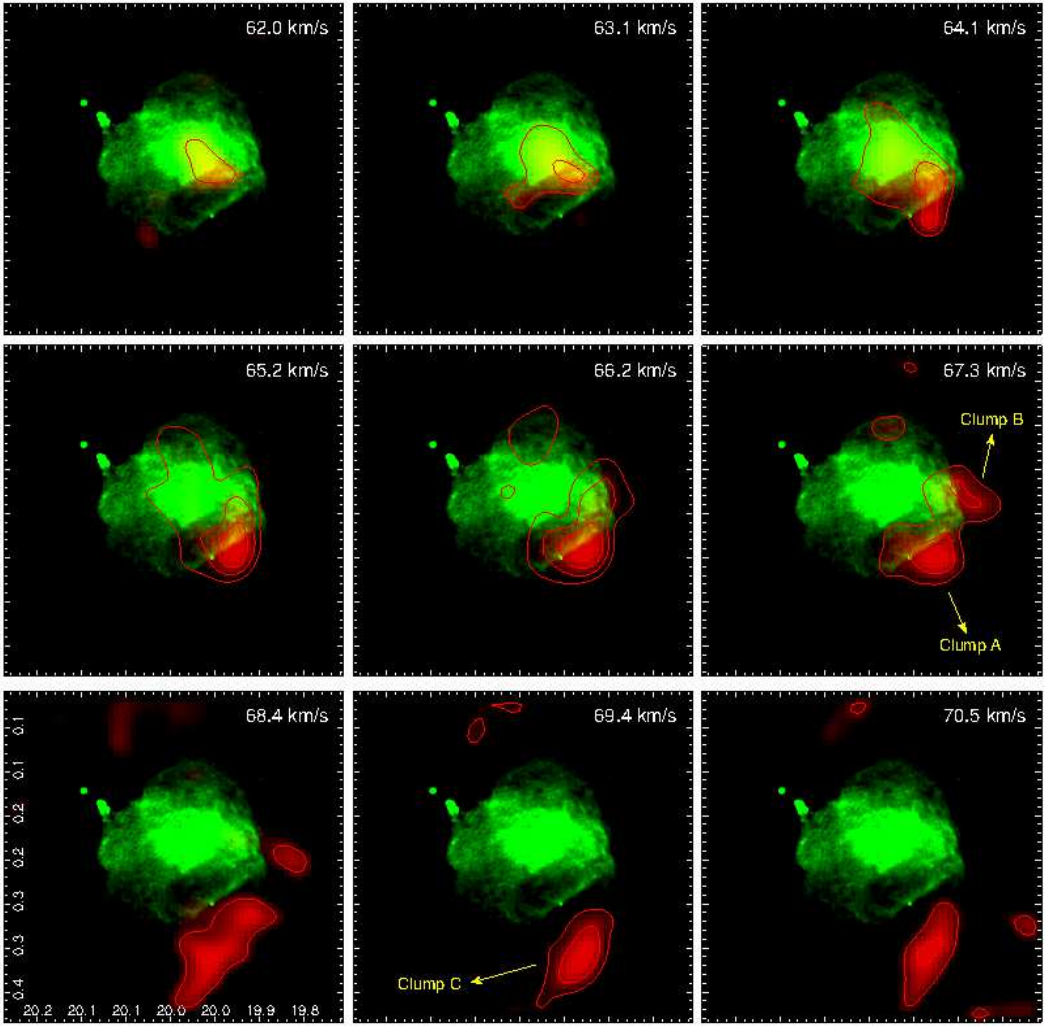}
\caption{Two-color image toward G20.0-0.2 in Galactic coordinates. 
In red with contours: emission of the \3 integrated every 1.05 \km~in the velocity range between 62 \km~and 71 \km~(contour 
levels are 2.7, 4.0, and 5.3 K \km). In green: radio continuum emission of G20 at 20 cm. We have marked the two molecular clumps A and B 
that form the molecular cloud probably interacting with the SNR. A third clump (C) appears in the direction of the 
flattest border of G20 but is not in contact with the remnant.}
\label{molec3}
\end{figure*}

By inspecting the whole $^{13}$CO cube, we found a molecular feature
extending from 62 to 71 km s$^{-1}$ as the most probable structure to be in contact with the SNR.
In Fig. \ref{molec3}, we present the $^{13}$CO emission integrated every 1.05 km $^{-1}$ along
this velocity range, displayed in Galactic coordinates. 
We noted the presence of a cloud that shows an arc-shape morphology and delineates the 
southern border of the SNR. This cloud is composed of two molecular clumps, labeled A and B.
Molecular clump A lies right upon the flattest border of G20 (refer to Fig. \ref{fig_radio}), while clump B is located
adjacent to the SE border of the SNR. A third molecular structure (labeled clump C) appears in the direction of 
the flattest border, but is not in contact with the radio emission.

\begin{figure}[h]
\centering
\includegraphics[width=6cm]{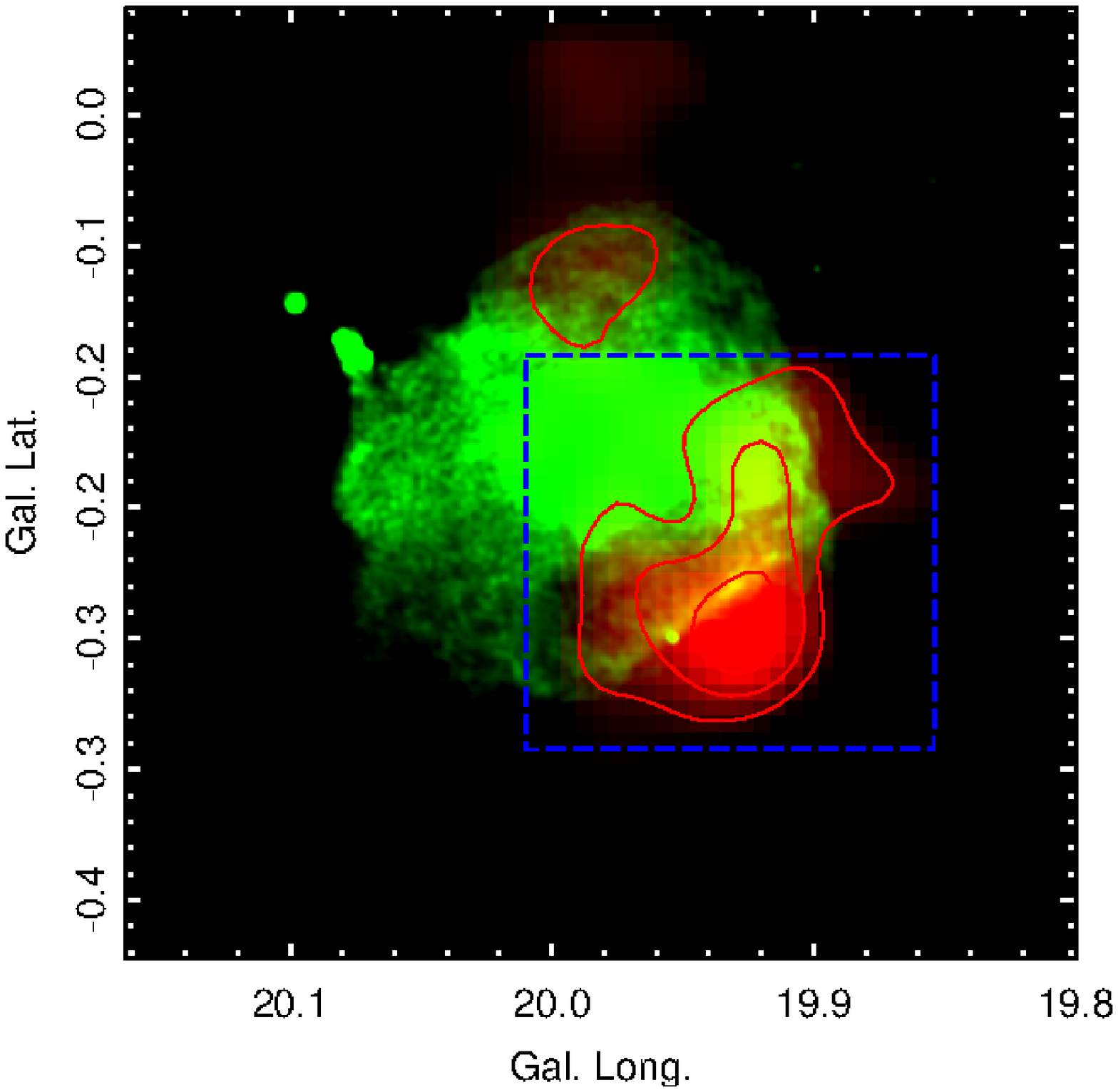}
\includegraphics[width=6cm]{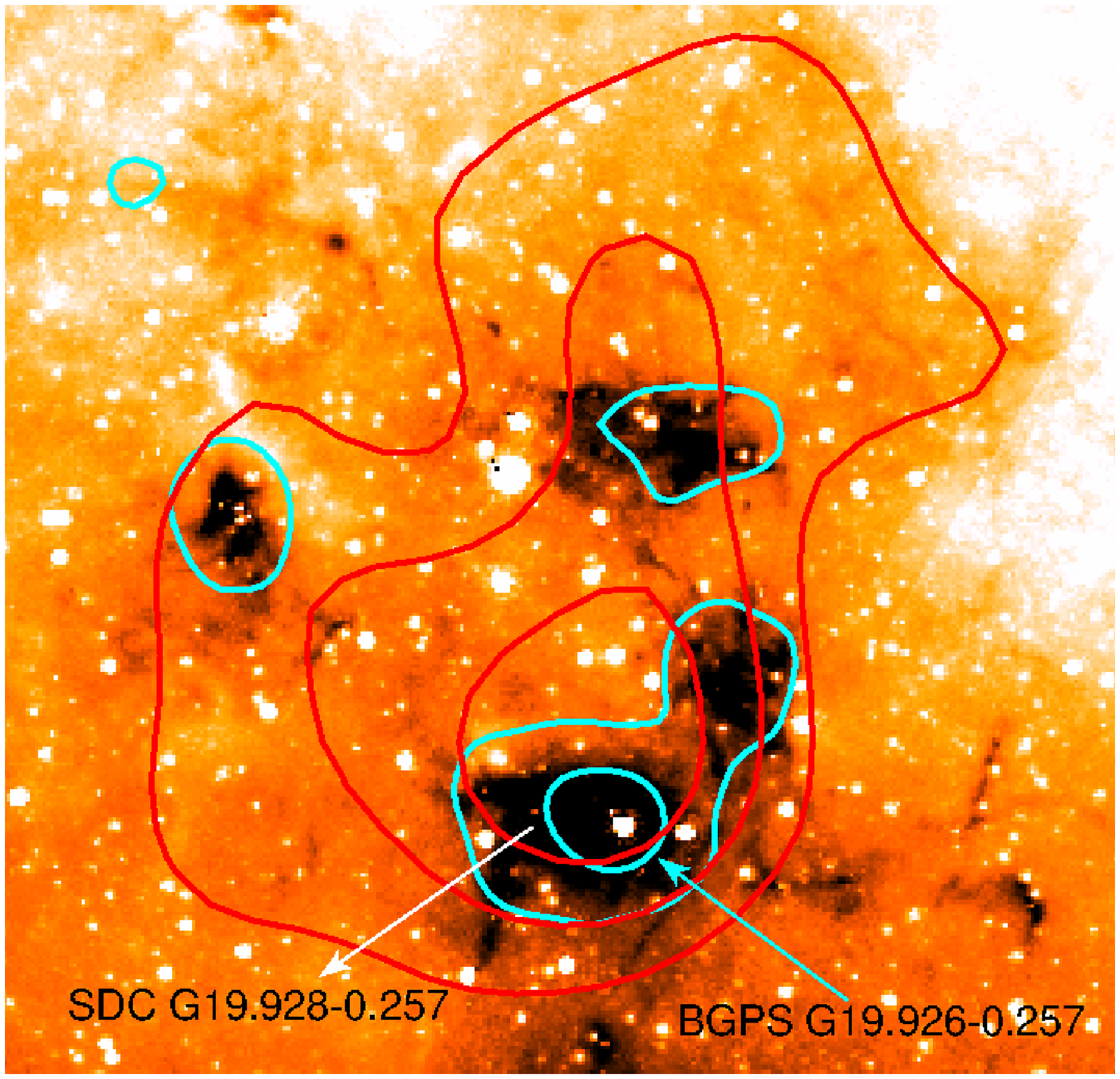}
\caption{{\it Left}: two-color image toward G20.0-0.2. In green: radio continuum emission at 20 cm. 
In red: emission of the \3 integrated between 64.3 and 68.6 \km~(contour levels are 10.0, 14.6, and 19.0 K \km). The blue
rectangle indicates the area shown in the adjacent figure.
{\it Right}: {\it Spitzer}-IRAC 8 $\mu$m band. The red contours are the \3 emission and the cyan contours are the 
1.1 mm continuum emission from the BGPS (contour levels are 0.1 and 0.5 Jy/Beam). The most intense molecular emission
(traced by the 19.0 K \km~contour level) shows a spatial correspondence with the millimeter 
source BGPS G19.926-0.257 and the dark cloud SDC G19.928-0.257.}
\label{clumpA}
\end{figure}

The molecular clump A spatially coincides with the Bolocam\footnote{The Bolocam Galactic Plane Survey (BGPS) is a 1.1 mm 
continuum survey of the Galactic Plane made using Bolocam on the Caltech Submillimeter Observatory, 
with a 33\pp~FWHM effective resolution \citep{aguirre11}.} millimeter continuum source \object{BGPS G19.926-0.257}.
Besides, this source lies within the ellipse of the Spitzer dark cloud \object{SDC G19.928-0.257} \citep{peretto09}.
In Fig. \ref{clumpA}, we show the excellent spatial correspondence among the molecular emission, the 
millimeter continuum source, and the dark cloud.

\begin{figure}[h]
\centering
\includegraphics[width=6cm,angle=-90]{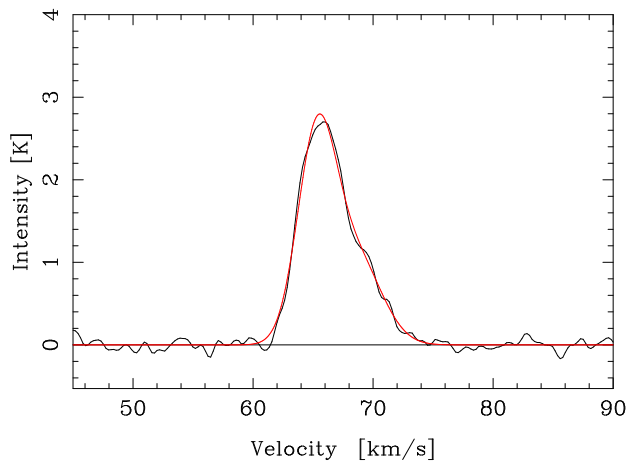}
\caption{Averaged spectrum obtained from a region toward the SNR flattest border,
centered at l=19.944\d, b=-0.245\d. The fit with two Gaussian functions is shown in red. }
\label{CO_spec}
\end{figure}

Analyzing the $^{13}$CO spectra toward the borders of the SNR, we found some interesting spectral features. 
Fig. \ref{CO_spec} displays an averaged spectrum obtained from a region toward the SNR's flattest border, centered at l=19.944\d, b=-0.245\d. 
The spectrum is not symmetric and presents a slight spectral shoulder toward higher velocities. 
The red curve in the figure is the result of the fitting with two Gaussian functions 
with central velocities and FWHM $\Delta$v of 65.3 and 68.5 km s$^{-1}$, and 3.8 and 5.0 km s$^{-1}$, respectively.
Such asymmetry could be evidence of turbulent motion in the gas, maybe produced by the SNR shock (see e.g. \citealt{falgarone94}), 
although we can not discard the presence of multiple molecular components.

Assuming that 66 km s$^{-1}$ is the central velocity of the cloud,
according to the Galactic model of \citet{fich89}, this cloud could be located at the distances of either
4.5 and 11.5 kpc. To resolve the ambiguity, we followed the procedure described in \citet{roman09}. Molecular 
clouds have HI embedded within them, which is cooler than the Galactic inter-cloud HI. 
Therefore, a MC located at the near distance absorbs the radiation from the warm HI at 
the far distance with the same velocity. As a consequence, the HI spectrum toward a MC at the near distance 
shows an absorption feature whose velocity coincides with the velocity of the molecular emission line.

\begin{figure}[ht]
\centering
\includegraphics[width=9cm]{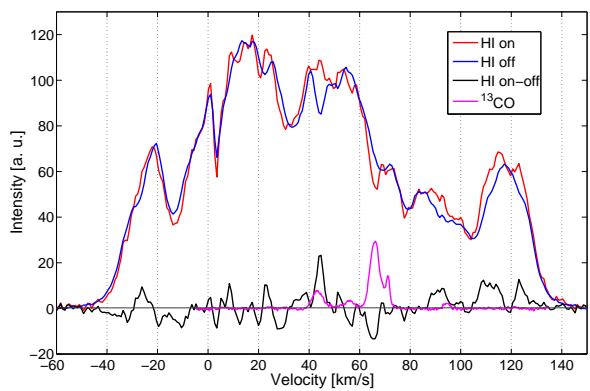}
\caption{\3 and HI spectra toward molecular clump A: \3 emission spectrum (magenta), HI on-position spectrum (red), HI off-position spectrum (blue), and HI absorption spectrum on-off (black).}
\label{HI}
\end{figure}

We constructed an HI absorption spectrum toward the clump A, which presents the most convincing 
signature of interaction with the SNR. 
The HI data were extracted from the VLA Galactic Plane Survey (VGPS, \citealt{stil06}), which maps the HI 21 cm line emission
with angular and spectral resolutions of 1$^{\prime}$ and 1.3~km~s$^{-1}$, respectively.
We obtained the \3 and HI on-position spectra from a rectangular region within the 19 K \km~contour level 
(see Fig. \ref{clumpA}, {\it left}) and outside the SNR to avoid the effect of the radio continuum source in the absorption spectrum. 
The off-position HI spectrum was made by averaging three HI spectrum from rectangular regions adjacent to clump A and free of \3 emission. 
The HI absorption spectrum results from the subtraction of the HI on- and off-position spectra. 
In Fig. \ref{HI}, we show the HI and \3 spectra toward the molecular clump A.
We noted the presence of an HI absorption line at a velocity of $\sim$65 \km, which coincides with a \3 emission line from the cloud.
From this analysis, we favor the near distance of 4.5 kpc for molecular clump A and thus for the SNR G20.0-0.2.

An additional evidence that the near kinematic distance could be the most likely one is 
the presence of the millimeter source BGPS 019.926-00.257 lying inside the Spitzer dark cloud SDC G19.928-0.257, both
of them associated with the molecular clump A (see Fig. \ref{clumpA}, {\it right}).
In fact, dark clouds are clouds of dust that appear dark as they absorb the mid-infrared radiation of the Galaxy. For this reason, 
they are usually placed at the near kinematic distance \citep{schlin11}.
Thus, we consider that the SNR (located at $\sim$4.5 kpc) and the UC HII region complex GAL 20.08-0.14 (at $\sim$12.6 kpc) are not related.

To estimate the mass and density of the molecular clumps, we assumed local thermodynamic equilibrium (LTE).
For the CO column density we used:
\begin{equation}
{\rm N(^{13}CO)} = 2.42 \times 10^{14} \frac{T_{\rm ex}+0.88}{1-e^{-5.29/T_{\rm ex}}} \int{\tau_{13} dv},
\label{eq1}
\end{equation}
where $T_{\rm ex}$ is the excitation temperature of the \3 transition and $\tau_{13}$ is the optical depth of the line.
Assuming that the \3 J=1--0 line is optically thin, we used the approximation
\begin{equation}
\int{\tau_{13} dv} \sim \frac{1}{J(T_{\rm ex}) - J(T_{\rm b})} \int{T_{\rm B} dv},
\label{eq2}
\end{equation}
where $J(T)=5.29(e^{5.29/T}-1)^{-1}$, $T_{b}=2.7$~K is the background temperature,
and $T_{B}$ is the brightness temperature of the line. 
To obtain the H$_2$ column density, we took the relative abundance from \citet{simon01}: 
N(H$_{2}$)/N(\3)$ \sim 5 \times 10^5$.
The mass of the molecular clumps was calculated from
\begin{equation}
{\rm M} = \mu~m_{{\rm H}} \sum{\left[ D^{2}~\Omega~{\rm N(H_{2})} \right] }, 
\label{eq3}
\end{equation}
where $\Omega$ is the solid angle subtended by the \3 J=1--0 beam size, $m_{\rm H}$ is the hydrogen mass,
$\mu$ is the mean molecular weight assumed to be 2.8 by taking into account a relative helium abundance
of 25 \%. Our summation was performed over the area of each molecular clump.
The column density was calculated assuming $T_{ex}=10$~K and the mass and number density considering 
a distance of 4.5 kpc. In Table \ref{table1}, we report the physical parameters of the molecular clumps A and B.
The obtained masses and densities are similar to those measured toward other molecular clouds interacting 
with SNRs (see \citealt{paron12,dubner04}).

\begin{table}[ht]
\caption{Physical characteristics of molecular clumps A and B: LSR central velocity (v$_{c}$), 
velocity width ($\Delta$v), size, H$_{2}$ column density, mass, and
number density. The linear size, mass and number density were calculated assuming a distance of 4.5 kpc. }
\label{table1}
%\tiny
\centering
\setlength{\extrarowheight}{3pt}
\begin{tabular}{ccccccc}
\hline\hline
Clump  & v$_{c}$  & $\Delta$v &    Size                 		 & N(H$_{2}$)          & Mass   &  Density  \\
       &   [\km]       &     [\km]    & [arcmin (pc)]             	 & [$\times10^{21}$ \cm2]   & [\msol]  &   [\ca]     \\ 
\hline
A     & $\sim$ 66  & 6   & 2.8$\times$3.8 (3.7$\times$5.0) &       9.9             &  12700 & 550     \\
B     & $\sim$ 69  & 4   & 4.3$\times$2.2 (5.6$\times$2.9) &       4.0             &  4500  & 230     \\
%C     & $\sim$ 70  & 5   & 4.4$\times$2.2 (5.8$\times$2.9) &       5.7             &  6400  & 320     \\
\hline
\end{tabular}
%\tablefoot{2MASS Qual.:.}
\end{table}

Additionally, we calculated the total hydrogen column density N(H), which is an important parameter
in the modelling of X-ray emission (see Sect. \ref{spectrum}).
For this purpose, we summed the contributions from both neutral 
and molecular hydrogen column densities: N(H) = N(HI)+2N(H$_{2}$). The integration was performed 
in the velocity range between 0 and 66 \km~and over the central core, in a region delimited by the 3.0 mJy/beam contour level 
of the radio continuum emission (see Fig. \ref{figX1a}), where the diffuse X-rays originate. 
The molecular column density N(H$_{2}$) was obtained from Eq. \ref{eq1} and the neutral column density N(HI) from 
the following equation: 
\begin{equation}
N(HI) = 1.82 \times10^{21} \int{T_{\rm B} dv}.
\label{eq4}
\end{equation}
We obtained N(H)~$\sim4\times10^{22}$ \cm2.      
   
\subsection{Young stellar objects embedded in the molecular gas}
\label{ysos}

The expansion of the shock front of a SNR into the ISM gives rise to a broad range of phenomena. 
SNRs can compress molecular clouds and induce the formation of inhomogeneities in the gas, which may collapse 
into dense molecular clumps where, eventually, new stars may form. 
Up to the present, there is no concluding observational evidences indicating that the SNRs can trigger star formation.
Only a few studies have revealed the existence of young stellar objects (YSOs) 
in the periphery of SNRs, such as the filled-center remnants G24.7+0.6 \citep{petriella10} 
and G54.1+0.3 \citep{koo08}, and the shell-type G59.5+0.1 \citep{xu12}, G0.1-0.1, G6.41-0.1 (W28), and G355.9-2.5 \citep{marquez-lugo10}. 
The main controversy regarding star formation triggered by a SNR is the difference between two timescales, namely the 
SNR's age ($\sim$10$^{5}$~yr) and the YSO's characteristic age ($\sim$10$^{6}$~yr), which would make 
it impossible to directly observe a SNR inducing the collapse of molecular cores to form new stars.
However, the progenitors of core-collapse SNs are massive stars, 
whose strong stellar wind during the duration of the main sequence phase ($\gtrsim$10$^{6}$~yr) can provide 
additional injection of energy into the molecular gas prior to the SNR.

\begin{figure}[h]
\centering
\includegraphics[width=9cm]{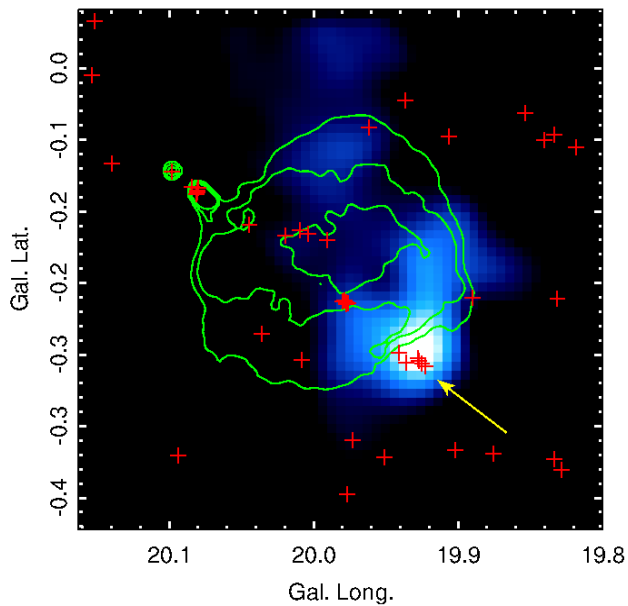}
\caption{Emission of the \3 integrated between 64.3 and 68.6 \km. The green contours are the radio continuum emission from G20.
The red crosses are the class I YSO candidates selected following the procedure described in the text. Several YSO candidates
appear grouped together in small clusters. One of them (indicated with an arrow) is located right upon the molecular 
gas probably shocked by the SNR.}
\label{fig_yso}
\end{figure}

In the case of G20, the radio and X-ray properties indicate that it is a core-collapse SNR. In addition, we
showed that the remnant is very likely interacting with a MC. 
So, it appears as a proper candidate to study star forming activity in its vicinity. 
To look for YSOs candidates around G20, we constructed a color-color (cc) diagram in the {\it Spitzer}-IRAC bands
[5.8]-[8.0] vs [3.6]-[4.5] (not shown here) with sources that report flux measurements in the four bands.
We applied the criterion of \citet{allen04} to identify class I YSO candidates, i.e. young stars at the earlier stages of evolution. 
In Fig. \ref{fig_yso}, we show the distribution of the class I YSO candidates, overimposed to the molecular cloud probably 
interacting with G20.
From this figure, we noted the presence of several clusters of YSO candidates, but we call special attention to a group of 6 YSO candidates that appear projected on the clump A, which is associated with the millimeter 
source BGPS G19.926-0.257 and with the dark cloud SDC G19.928-0.257. 
The presence of dense molecular material, a dark cloud, abundant dust (traced by the millimeter source), and YSO candidates 
suggests that this region could be an active star forming site \citep{rath07,rath06},
representing another potential site of star formation near a SNR.

\section{Concluding remarks}
\label{secc_summ}

We have presented the first X-ray study of the SNR G20.0-0.2 using {\it Chandra} observations. 
We detected diffuse emission, which
has a very good correlation with the bright central radio emission. The X-ray spectrum can be equally well fitted
by a thermal plasma in a non equilibrium or a power-law model. Taking into account the
obtained photon index $\Gamma\sim1.9$ and the radio properties of the SNR, we conclude 
that a non-thermal origin of the X-ray emission is the most probable. 
In addition, we have reported the presence of the hard X-ray point source CXO J182807.4-113516 located at the geometrical center of the remnant, which is characterized by a non-thermal spectrum and a X-ray luminosity between $\sim0.7-1.2\times10^{32}$ erg/s in the 0.5-8.0 keV band (similar to young pulsars), suggesting that it is a good candidate to be the central compact source powering the PWN.

The study of the surrounding interstellar gas around G20 allowed us to constrain its distance to a value of 4.5 kpc.
We found a molecular cloud with some indications of interaction with the flattest border of the radio emission, 
and in excellent spatial correspondence with the millimeter continuum source BGPS G19.926-0.257 and the dark cloud 
SDC G19.928-0.257. We identified a group of class I YSO candidates located in the brightest region of the molecular cloud.
The presence of these young sources, a dark cloud and abundant dust suggest that the region around G20 could
be a potential star forming site. 

\begin{acknowledgements}
A.P. is a doctoral fellow of CONICET, Argentina. S.P. and E.G. are members of 
the {\sl Carrera del investigador cient\'\i fico} of CONICET, Argentina.
This research was partially supported by Argentina Grants awarded by CONICET, ANPCYT and 
UBACYT 20020100100011. 
\end{acknowledgements}

%%%%%%%%%%%%%%%%%%%%%%%%%%%%%%%%%%%%%%%%%%%%%%%%%%%%%%%%%%%%%%%%%%%%%
\bibliographystyle{aa}  % A&A format
   %\bibliographystyle{klunamed}     
   % format of references provided by the review (.bst)
\bibliography{biblio}
   % file containing the bibtex references (.bib)
\IfFileExists{\jobname.bbl}{}
{\typeout{}
\typeout{****************************************************}
\typeout{****************************************************}
\typeout{** Please run "bibtex \jobname" to obtain}
\typeout{** the bibliography and then re-run LaTeX}
\typeout{** twice to fix the references!}
\typeout{****************************************************}
\typeout{****************************************************}
\typeout{}
}

\label{lastpage}

\end{document}